# Turbulent Couette Flow:
# An analytical solution

Trinh, Khanh Tuoc

Institute of Food Nutrition and Human Health

Massey University, New Zealand

*K.T.Trinh@massey.ac.nz*

## Abstract

A simple analytical solution for turbulent plane Couette flow is obtained from a subset of the Navier-Stokes equations. This approach analyses the effect of the unsteady state Lagrangian diffusion of viscous momentum on the smoothed phase velocity. Three initial velocity profiles at the start of a turbulent wall cycle are considered. The predictions adequate identify the existence of three layers usually observed in turbulent flow fields, in particular a viscosity dominated wall layer and a log-law layer. Experimental measurements of the time-averaged velocity are also well correlated.

## 1    Introduction

In previous postings (Trinh, 2009c, Trinh, 2010e) we have seen that the Stokes solution for a plate suddenly in motion (Stokes, 1851) can accurately correlate the time-averaged velocity profile in the wall layer of turbulent flows as first proposed by Einstein and Li (1956) and others (Hanratty, 1956, Meek and Baer, 1970). In fact the Stokes solution is able to reproduce many of the classical statistics on turbulence including the normalised fluctuations velocity and the probability density distribution of the instantaneous velocity (Trinh, 2010c). This approach highlighted the role viscous diffusion of momentum in turbulence and allowed us to construct a very simple visualisation of turbulent flow fields based on the coexistence of several regions that follow very different mechanisms (Trinh, 2010a). In particular, it allowed us to deal effectively with the dissipation scales and develop a procedure for a theoretical closure of the Reynolds equations (Trinh, 2010d). There are two important differences in physical visualisation between the present theoretical construction and previous approaches.

## 1.1 The physical background

Firstly the instantaneous velocity is decomposed into 4 components (Trinh, 2009a) not two as in Reynolds' analysis . A typical trace of the instantaneous velocity near the wall is shown in Figure 1.

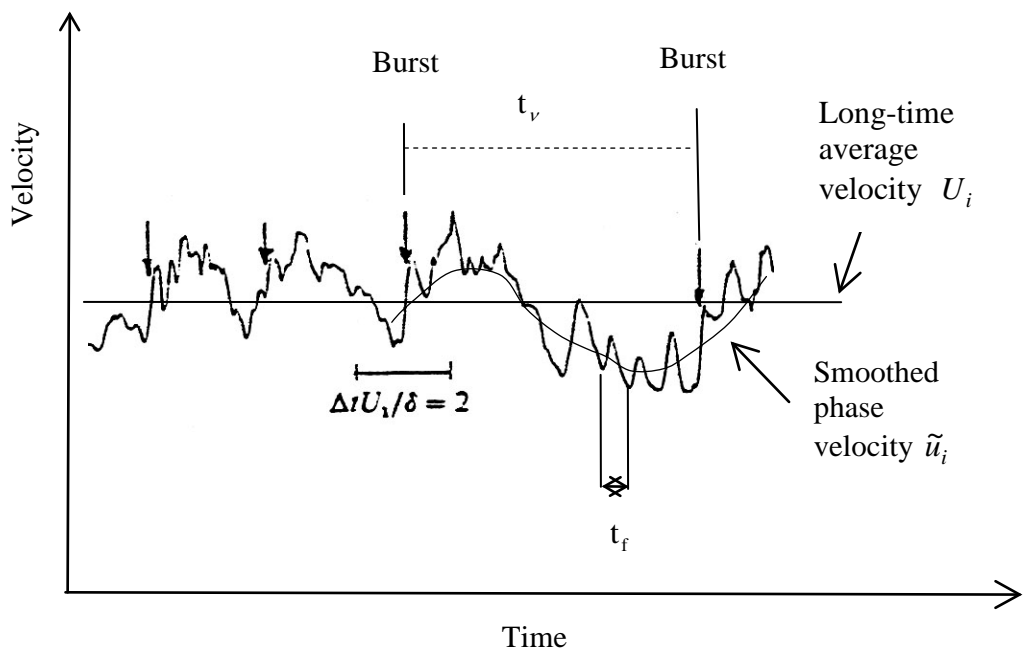

Figure 1 Trace of instantaneous streamwise velocity near the wall reproduced from (Trinh, 2009c) after measurements by Antonia et al. (1990)

This velocity trace reflects the classical wall cycle process first identified by Kline et al. (1967) through hydrogen bubble visualisation. It starts with an inrush of fluid from the main flow towards the wall followed by a sweep phase characterised by a vortex travelling above the wall. The vortex induces a sub-boundary layer flow underneath its path, seen in the bubble visualisation as a streak of low speed fluid. The low-speed streak grows and oscillates until the wall fluid is ejected towards the outer region in a violent burst (Figure 2).

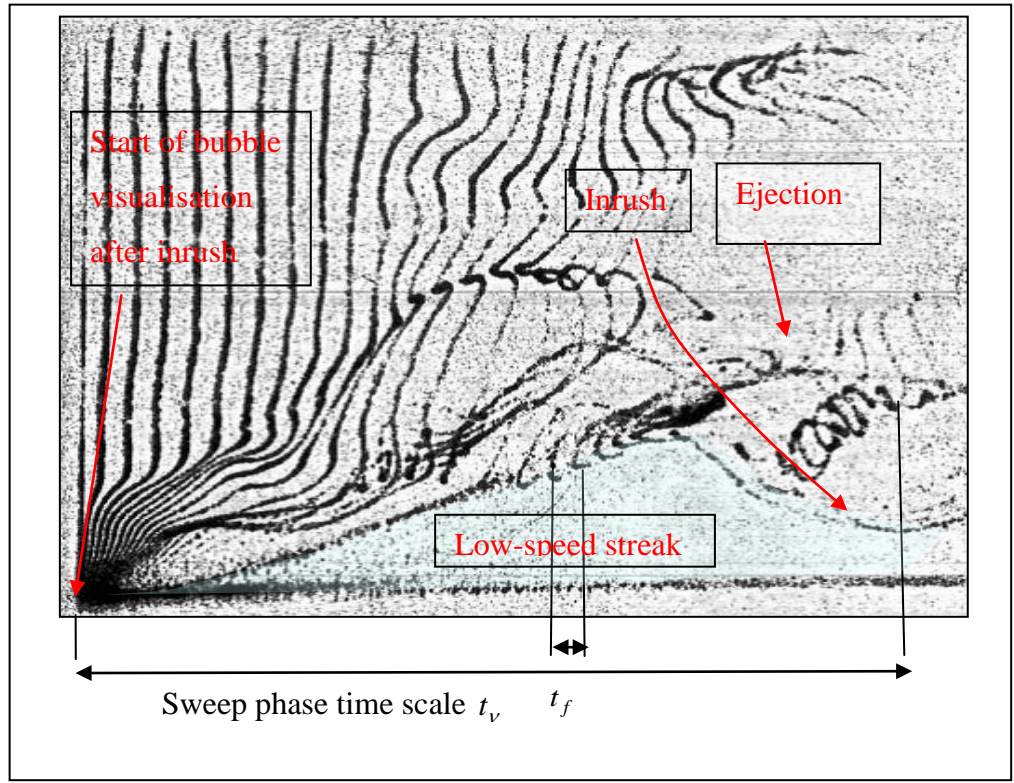

Figure 2  Velocities and time scales associated with the low speed streaks and vortices in the wall layer. Photograph adapted from (Kim et al., 1971).

Reynolds (1895) decomposed the instantaneous velocity into two components

$$u_i = U_i + U'_i \qquad (1)$$

with

$$U_i = \lim_{t \to \infty} \int_0^t u_i \, dt \qquad (2)$$

$$\int_0^\infty U'_i \, dt = 0 \qquad (3)$$

The instantaneous velocity may be decomposed as:

$$u_i = \tilde{u}_i + u'_i \qquad (4)$$

where $\tilde{u}_i$ is the smoothed phase velocity within the coherent structure. The fast fluctuations $u'_i$ typically imposed by travelling vortices, and highlighted by the bubble visualisation in Figure 2, may be assumed to be periodic with a timescale $t_f$ and therefore do not contribute to the long term average. Hence

$$U'_i = \tilde{U}'_i + u'_i \qquad (5)$$

$$\tilde{U}'_i = \tilde{u}_i - U_i \qquad (6)$$

Then

$$u_i = U_i + \tilde{U}'_i + u'_i \qquad (7)$$

The interaction of convected inertial effects of forced oscillations with viscous effects near a wall results in a non-oscillating motion that is referred to in the literature as "Streaming" e.g. (Tetlionis, 1981, Schlichting, 1960) and lead to a further velocity term that must be included in equation (7)

$$u_i = U_i + \tilde{U}'_i + u'_i(\omega t) + u_{i,st} \qquad (8)$$

This four component decomposition leads to the identification of two types of Reynolds stresses: slow and fast. The fast Reynolds stresses arise from the streaming flow, also called transient shear growth by Schoppa and Hussain (2002). In this construction, the ejections form the defining difference between unsteady laminar flow and turbulent flow. Solutions such as that of Stokes neglect the fast fluctuations and thus cannot account for the critical fast Reynolds stresses. In the present construction these solutions are used to model the smoothed phase velocities $\tilde{u}_i$ that are found in the coherent structures, most importantly in the sub-layers defined by the low speed streaks at the wall (Trinh, 2009c, Trinh, 2009a).

Secondly the neglect of the convection terms in the Navier-Stokes equations for a flat plate

$$\frac{\partial u}{\partial t} + u\frac{\partial u}{\partial x} + v\frac{\partial u}{\partial y} = \nu\frac{\partial^2 u}{\partial y^2} \qquad (9)$$

to obtain the governing equation in the Stokes solution

$$\frac{\mathcal{D}\tilde{u}}{\mathcal{D}t} = \nu\frac{\mathcal{D}^2\tilde{u}}{\mathcal{D}y^2} \qquad (10)$$

cannot be justified in an Eulerian context because the terms $\partial u/\partial t$, $u\partial u/\partial x$ and $v\, \partial u/\partial y$ are of the same order of magnitude. The argument is best made in the context of a new Eulerian partial derivative $\mathcal{D}/\mathcal{D}t$ following the path of diffusion (Trinh, 2010b) that allows us to decouple the convection and diffusion terms.

In this paper we use this new partial derivative to obtain an analytical solution for turbulent Couette flow.

## 1.2  Couette flow

This classic problem of a fluid sheared between two parallel plates separated by a distance H has interested many investigators since the study of Couette (1890). The greatest attraction is simplicity in the flow pattern that has a linear velocity profile in laminar flow. This was the basis to explain Newton's law of viscosity. Experimental measurements that approximate real Couette flow between parallel plates are made with concentric cylinders starting with the classic work of Taylor (1923, 1936a, 1936b). Following the investigation of Coles (1965) the flow between rotating cylinders has been used to investigate the transition from laminar to turbulent fluid motion and over 2000 experimental, numerical and theoretical studies have been performed for this geometry (Caton et al., 2000) but relatively few measurements of turbulent velocity profiles exist notably by Reichardt (1956), Robertson and Johnson (1970), Geach (1970), El Telbany and Reynolds (1982), Aydin and Leutheusser (1991) and Bech et al. (1995). Dorfman (1963) reported that velocity profiles of Reichardt followed the well-known log-law. Fehrenbacher et al (2007) have given a more detailed experimental description of structures and statistics in turbulent Couette flow.

Measurements of truly plane Couette flow were made by Kitoh et al. (2005) with a conveyor belt moving in a 0.88m by 5.12m channel in the resurgence in interest in turbulent Couette flow of the last ten fifteen years. They confirm that the time averaged velocity profile can be divided four layers usually observed in other turbulent flows: a laminar sub-layer, a buffer, a log-law layer and a velocity defect layers. But they report that the log-law layer in Couette flow is 2-2 times larger than in Poiseuille flow. The same authors (Nakabayashi et al., 2004) noted that the defect law in Poiseuille and Couette flow are quite different. Kitoh et al. also provided the first comprehensive statistics in truly experimental plane Couette flow. They further identify large longitudinal vortices not seen in pipe flow that extend the entire height of the channel and cause a slow fluctuation with large amplitude in the streamwise velocity component. These vortices had been identified in the DNS of Lee and Kim (1991), Komminaho et al. (1996)and Papavassiliou and Hanratty (1997). Many other workers also use numerical methods to investigate Turbulent Couette flow. Andersson and Pettersson (1994), Heinloo (2004), used CFD with k-$\varepsilon$ and other closure techniques. Kawahara and Kida (2001) illustrated by DNS the sweep-burst cycle with a streaky pattern of low-speed streaks similarly to that shown in Figure 2. Bilson and Bremhorst (2007) illustrated the influence of near-wall vortex pairs, Itano and Generalis (2009) used a Hairpin Vortex Solution to analyse the flow pattern, Viswanath (2007) used five different solutions to demonstrate the recurrent breakup and re-formation of near-wall coherent structures in turbulent Couette flow, Duguet et al. (2010)analysed turbulent patterns near transition, Barkley and Tuckerman (2007) studied specifically a turbulent–laminar banded pattern in plane Couette flow photographed in Taylor-Couette flow (Prigent et al., 2002).

In other approaches Eigolf and Weiss (1995) used a difference-quotient turbulence model, Moehlis et al. (2002) used proper orthogonal decomposition. A number of authors referred to the theory of strange attractors e.g. (Brandstater and Swinney, 1987a, Brandstater and Swinney, 1987b) and their work is well reviewed by Gibson et al. (2008).

The friction factor in Plane Couette flow has been modelled empirically by Robertson and Johnson (1970) and confirmed by Kitoh et al.(op.cit.). In Taylor-Couette van den Berg et al. (2003) showed that the friction factor could be correlated with both logarithmic and power law relationships. Eckhardt et al. (2007b) noted that many authors, notably Lathrop et al.(1992) show further a non-trivial dependence of the dimensionless torque on the radius ratio and on the non-dimensional gap width. Eckhardt et al. and Delfos et al. (2009) studied turbulent Couette flow with both cylinders rotating. Eckhardt et al. approached the problem theoretically by using an analogy between Taylor-Couette flow and Rayleigh-Bernard convection.

## 2   Start-up Couette flow

Consider the situation when the upper plate is suddenly moved from rest with velocity $V$ at time $t = 0$. This system is treated in classic texts as start-up of laminar Couette flow e.g. (Bird et al., 1960, Theodore, 1971) and we start again with equation (10)

$$\frac{\mathcal{D}\tilde{u}}{\mathcal{D}t} = \nu \frac{\mathcal{D}^2\tilde{u}}{\mathcal{D}y^2}$$

We define the dimensionless velocity, distance and time as

$$\phi = \frac{\tilde{u}}{V} \qquad (11)$$

$$\xi = \frac{y}{H} \qquad (12)$$

$$\tau = \frac{\nu t}{H^2} \qquad (13)$$

Equation (2) becomes

$$\frac{\mathcal{D}\phi}{\mathcal{D}\tau} = \frac{\mathcal{D}^2\phi}{\mathcal{D}\xi^2} \qquad (14)$$

which is solved for the conditions

IC         $\tau = 0$         $0 \leq \xi \leq 1$         $\phi = 0$

BC1       $\tau > 0$         $\xi = 0$         $\phi = 0$

BC2       $\tau > 0$         $\xi = 1$         $\phi = 1$

We seek a solution of the form

$$\phi(\xi,\tau) = \phi_\infty(\xi) - \phi_t(\xi,\tau) \qquad (15)$$

At time $t \to \infty$, $\mathcal{D}\phi/\mathcal{D}\tau = 0$ and we obtain the familiar laminar Couette flow with a linear velocity profile

$$\phi_\infty = \xi \qquad (16)$$

Substituting into equation (15) gives

$$\frac{\mathcal{D}\phi_t}{\mathcal{D}\tau} = \frac{\mathcal{D}^2\phi_t}{\mathcal{D}\xi^2} \qquad (17)$$

We now use the method of separation of variables, setting

$$\phi_t = \Xi(\xi)T(\tau) \tag{18}$$

Substituting into (17), rearranging and equating both sides of the equation to a parameter $(-\alpha^2)$

$$\frac{1}{T}\frac{dT}{d\tau} = \frac{1}{\Xi}\frac{d^2\Xi}{d\xi^2} = -\alpha^2 \tag{19}$$

Equation (19) then gives

$$T = C_0 e^{-\alpha^2 \tau} \tag{20}$$

$$\Xi = (C_1 \sin \alpha\xi + C_2 \cos \alpha\xi) \tag{21}$$

$$\phi = \xi + C_0 e^{-\alpha^2 \tau}(C_1 \sin \alpha\xi + C_2 \cos \alpha\xi) \tag{22}$$

The constants $C_0, C_1, C_2$ are determined from the IC and BC.

BC1 gives $C_0 C_2 = 0$. Then BC2 gives

$$C_0 C_1 e^{-\alpha^2 \tau} \sin \alpha = 0 \tag{23}$$

which is satisfied if either $C_0 C_1 = 0$ or $\sin \alpha = 0$. Since $C_0 C_1 \neq 0$ because $\phi$ is a function of $\tau$, we put

$$\alpha = n\pi \quad \text{where} \quad n = 1,2,3,4... \tag{24}$$

and

$$\phi = \xi + \sum_1^\infty C_n e^{-\alpha^2 \tau} \sin(n\pi\xi) \tag{25}$$

where $C_n = C_0 C_1$ is determined by applying the IC then by multiplying both sides by $\sin(m\pi\xi)d\xi$ and integrating from 0 to 1. At $\tau = 0$ equation (25) gives

$$-\int_0^1 \xi \sin(m)d\xi = C_n \int_0^1 \sin(n\pi\xi)\sin(m\pi\xi)d\xi \tag{26}$$

Making use of standard integrals of trigonometric functions (Wikipedia, 2010)

$$-\int_0^1 \xi \sin(m\pi\xi)d\xi = -\left[\frac{\sin(m\pi\xi)}{m\pi}\right]_0^1 + \left[\frac{\xi\cos(m\pi\xi)}{m\pi}\right]_0^1 = \frac{(-1)^n}{m\pi} \tag{27}$$

Because of orthogonality, the right hand side only contributes when $m = n$

$$\int_0^1 \sin^2(n\pi\xi)d\xi = \left[\frac{\xi}{2}\right]_0^1 - \left[\frac{1}{2n\pi}\sin(n\pi\xi)\cos(n\pi\xi)\right]_0^1 = \frac{1}{2} \tag{28}$$

Then

$$C_n = \frac{2}{n\pi}(-1)^n \qquad (29)$$

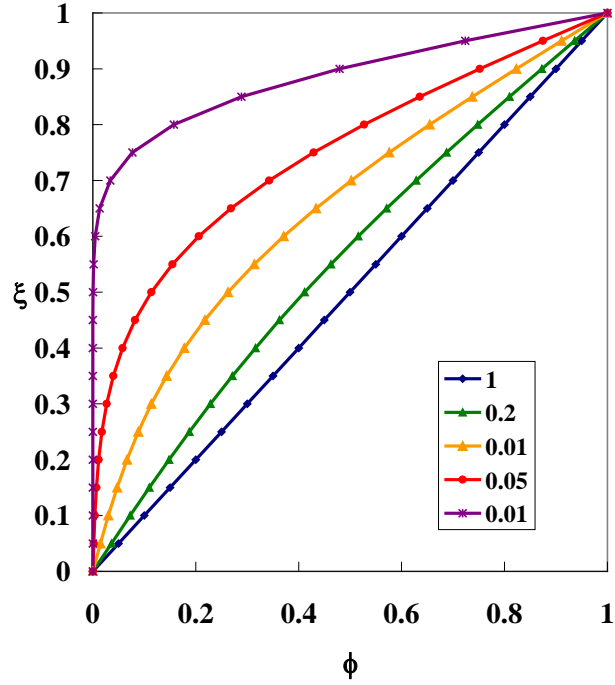

Figure 3 Evolution of velocity profile during start up Couette flow. Numbers indicate values of normalised time $\tau$.

Figure 3 shows the evolution of the instantaneous velocity profile with normalised time.

## 3 Turbulent Couette flow

We use equation (14) to model turbulent Couette flow by changing the IC to comply with the classic observations of the wall layer process first introduced by Kline et al. (1967), confirmed and further detailed by many others e.g. (Offen and Kline, 1974, Corino and Brodkey, 1969). This picture, which is discussed in more detailed in a previous overview of the full theoretical construction of a model for turbulence (Trinh, 2009c), shows that the diffusion of momentum expressed by equation (14) simplify modifies an original velocity profile found at time $\tau = 0$ at the inrush phase. If sufficient time is allowed then at $\tau = \infty$ we obtain equation (16) which represents therefore always a stable possible outcome of the solution. We thus need to modify the initial and boundary conditions for this exercise as

| | | | |
|---|---|---|---|
| IC | $\tau = 0$ | $0 \leq \xi \leq 1$ | $\phi = \phi_0$ |
| BC1 | $\tau > 0$ | $\xi = 0$ | $\phi = 0$ |
| BC2 | $\tau > 0$ | $\xi = 1$ | $\phi = 1$ |

where $\phi_0$ is the initial velocity profile yet to be stipulated. We will now do this in two steps.

## 3.1 Initial plug flow

For our first attempt we assume plug flow at the beginning of a cycle of the wall process

$$\phi_0 = \frac{1}{2} \tag{30}$$

Then at $\tau = 0$ equation (25) gives

$$\frac{1}{2} - \xi = \sum_1^\infty C_n \sin(n\pi\xi) \tag{31}$$

Multiplying both sides by $\sin(n\pi\xi)$ and integrating for $0 \leq \xi \leq 1$ gives

$$\int_0^1 \frac{1}{2} \sin(m\pi\xi) d\xi - \int_0^1 \xi \sin(m\pi\xi) d\xi = \int_0^1 \sum_1^\infty C_n \sin(m\pi\xi) \sin(n\pi\xi) d\xi \tag{32}$$

The new term is

$$\int_0^1 \frac{\sin(m\pi\xi)}{2} d\xi = -\left[\frac{\cos(m\pi\xi)}{2m\pi}\right]_0^1 = \frac{1}{2m\pi} - \frac{\cos(m\pi)}{2m\pi} \tag{33}$$

Then

$$C_n = \frac{1}{n\pi}\left[(-1)^n + 1\right] \tag{34}$$

Thus the instantaneous profile is

$$\phi = \xi + \sum_1^\infty \frac{\left[(-1)^n + 1\right]}{n\pi} e^{-\alpha^2 \tau} \sin(n\pi\xi) \tag{35}$$

Averaging over $\tau$

$$\Phi = \xi + \sum_1^\infty \frac{\left[(-1)^n + 1\right]}{\tau(n\pi)^3}\left[e^{-\alpha^2 \tau} - 1\right]\sin(n\pi\xi) \tag{36}$$

The plot of equation (36) against the measurements of Reichardt (1956) in Figure 4 shows that it is able to capture the essential form of the velocity profile. At $\tau = \infty$ the correspondence is naturally exact and the best prediction for the region near the walls occurs at $\tau = 0.025$ but the core region near the mid-point is not well represented. It is improved for $\tau = 0.035$ but then the wall region is less well correlated.

The time scale cannot of course be taken arbitrarily. Equation (13) can be rearranged as

$$\tau = \frac{\nu t}{H^2} = \frac{8T^{+2}}{f\,\mathrm{Re}^2} \qquad (37)$$

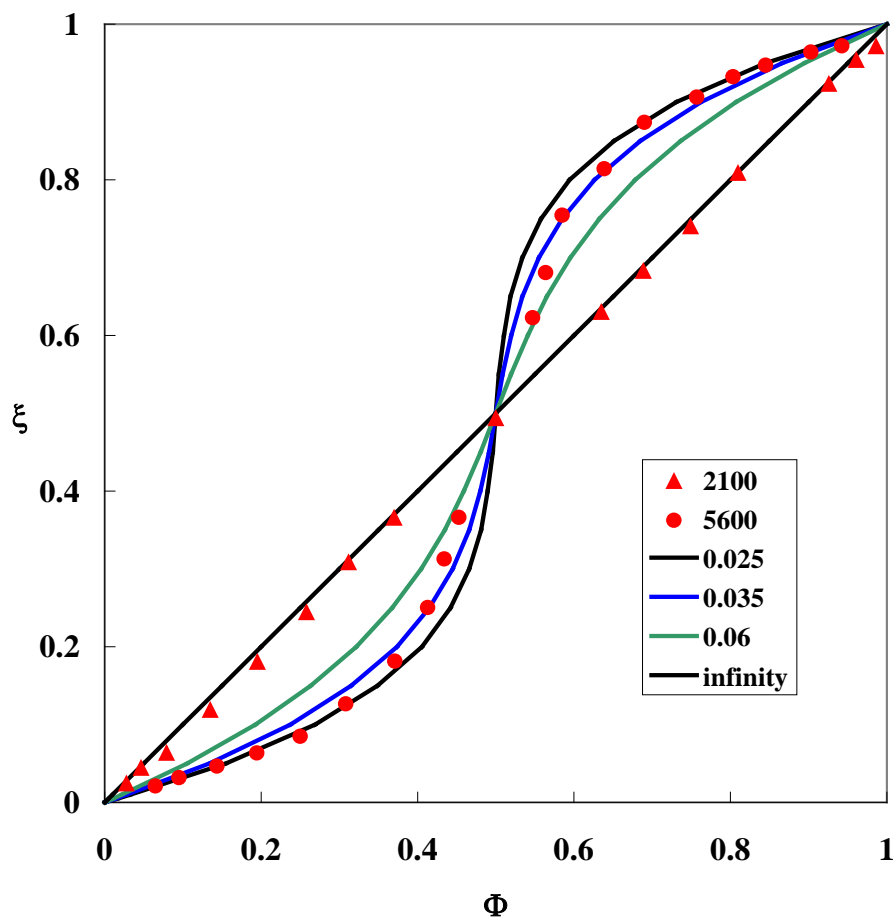

Figure 4. Comparison between profiles from equation (36) with the data of Reichardt (1956). Numbers refer to the value of the dimensionless time $\tau$.

where

$$T^+ = u_* \sqrt{\frac{t}{\nu}} \tag{38}$$

is the normalised time-scale used by Einstein and Li (1956), Meek and Baer (1970, 1973, 1973) and others,

$$\text{Re} = \frac{HV}{\nu} \tag{39}$$

$$f = \frac{2\tau_w}{\rho V^2} \tag{40}$$

$$u_* = \sqrt{\frac{\tau_w}{\rho}} \tag{41}$$

are the Reynolds number, friction factor and friction velocity and $\tau_w$ is the wall shear stress, not to be confused here with the normalised time scale $\tau$.

For turbulent flows, we have consistently used for many applications the value of $T^+ = 17.5$ that is constant with Reynolds number since the thickness of the wall layer itself is constant (Trinh, 2009c, Trinh, 2010a). The friction factor for turbulent flow between rotating cylinders can be estimated from a correlation presented by Dorfman (1963)

$$\frac{1}{\sqrt{f}} = 8.132 \log\left(\text{Re}\sqrt{\frac{f}{2}}\right) + 3.493 \tag{42}$$

While this formula does not account for the ratio of the cylinder diameters that is known to affect the friction factor e.g. (Eckhardt et al., 2007b) it is a good approximation for turbulent Couette flow where this ratio is close to unity. For $\text{Re} = 5800$, equation (37) gives $\tau \approx 0.019$ of the same order of magnitude as the best fit value of $\tau = 0.025$ in Figure 4.

Comparison between Figures (3) and (4) shows that the substitution of initial plug flow for the initial zero velocity profile has dramatically shifted the prediction to a system of two symmetric profiles. An even more interesting consequence is observed when the velocity profile is plotted in normalised coordinates $y^+ = yu_*/\nu$ and $U^+ = U/u_*$ where $U$ is the time averaged value of $\tilde{u}$.

Figure 5 shows that equation (36) predicts the existence of a small zone near the wall where viscous diffusion predominates similar to the assumption of a laminar sub-layer by Prandtl (1935) and an adjacent log-law region. This was achieved without any assumption concerning the structures that populate turbulent flow and confirm the similarity argument of Millikan (1938) that I have supported for many years. Essentially the transition from a region that scale with wall parameters $u_*, \nu$ and a region that scale with the outer parameters $V, H$ must obey a log-law.

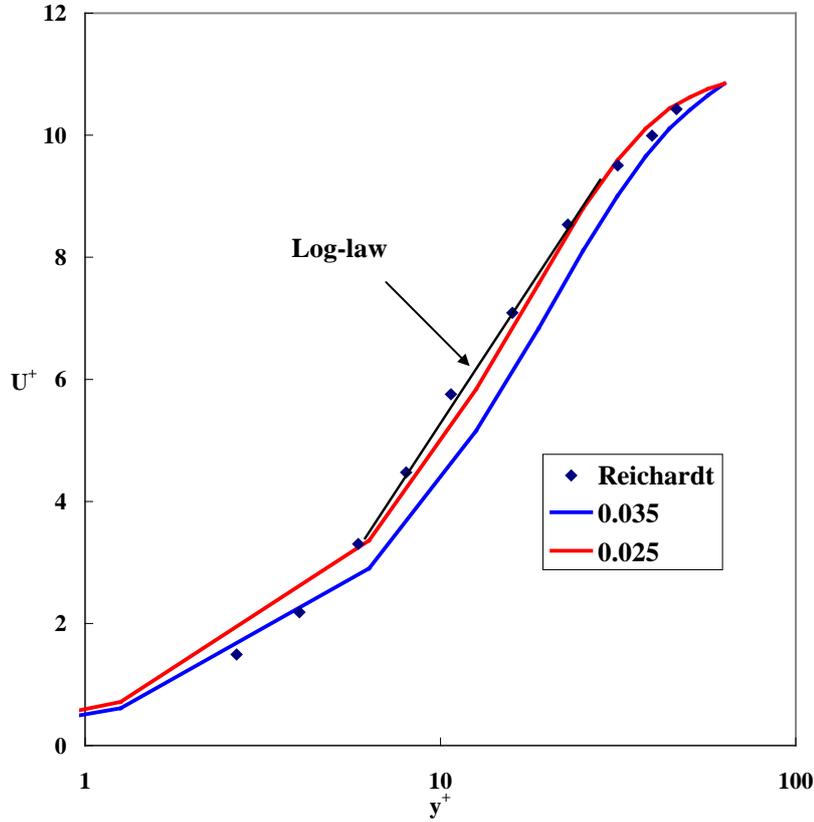

Figure 5. Log law representation of the velocity profiles measured by Reichardt (1956) and predicted by equation (36).

Despite the encouraging results achieved with this simple approach, we cannot ignore the fact that the core region is not well modelled by equation (36) and we need therefore to look at other postulations for the function $\phi_0$.

## 3.2 Initial full turbulence

The wall layer process begins with the inrush of fluid from the mainstream which is a direct consequence of the depletion created by the bursts or ejections of wall fluid. In fact the inrush and bursting phases overlap. Thus the initial profile in our analysis cannot be plug flow but a situation when turbulence reaches to the wall itself. Physically this means that we postulate that the wall and law-of-the-wake layers are extremely thin in the conditions of full turbulence created by the ejections and can be neglected. Figure 6 illustrates the initial velocity profiles that have been tested.

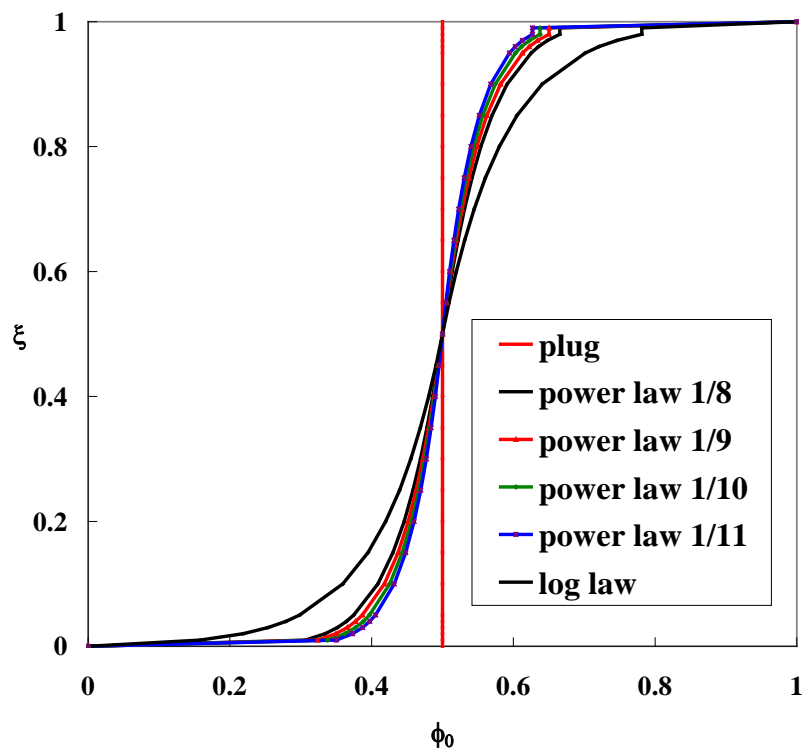

Figure 6. Possible models of initial velocity profiles

### 3.2.1 Initial power law profile

We first model the initial profile with a power law That applies to each half of the channel

$$\frac{\phi_0}{0.5} = \left(\frac{y}{(H/2)}\right)^p = \left(\frac{\xi}{0.5}\right)^p \qquad \xi \leq 0.5 \qquad (43)$$

Nikuradse (Nikuradse, 1932, Schlichting, 1960) showed that the exponent $p$ decreased with increasing Reynolds number. We expect that our assumption of complete turbulence would link with a very small value of $p$. In fact variations of the exponent p between $1/8 - 1/11$ have a relatively minor effect on the $\phi_0$ profile as shown in Figure 6.

The parameter $C_n$ in equation (29) can again be computed by performing the integral

$$\int_0^1 \phi_0 \sin(m\pi\xi)d\xi - \int_0^1 \xi \sin(m\pi\xi)d\xi = \int_0^1 \sum_1^\infty C_n \sin(m\pi\xi)\sin(n\pi\xi)d\xi \qquad (44)$$

giving

$$C_n = \frac{2(-1)^n}{n\pi} + 2C_4 \qquad (45)$$

where

$$C_4 = \int_0^1 \phi_0 \sin(m\pi\xi)d\xi \qquad (46)$$

Then equation (29) becomes

$$\phi = \xi + \sum_1^\infty \left[\frac{2(-1)^n}{n\pi} + 2C_4\right] e^{-\alpha^2\tau} \sin(n\pi\xi) \qquad (47)$$

Time averaging equation (47) gives

$$\Phi = \xi + \sum_1^\infty \left[\frac{2(-1)^n}{n\pi} + 2C_4\right] \frac{\left(e^{-n^2\pi^2\tau} - 1\right)}{n^2\pi^2\tau} \sin(n\pi\xi) \qquad (48)$$

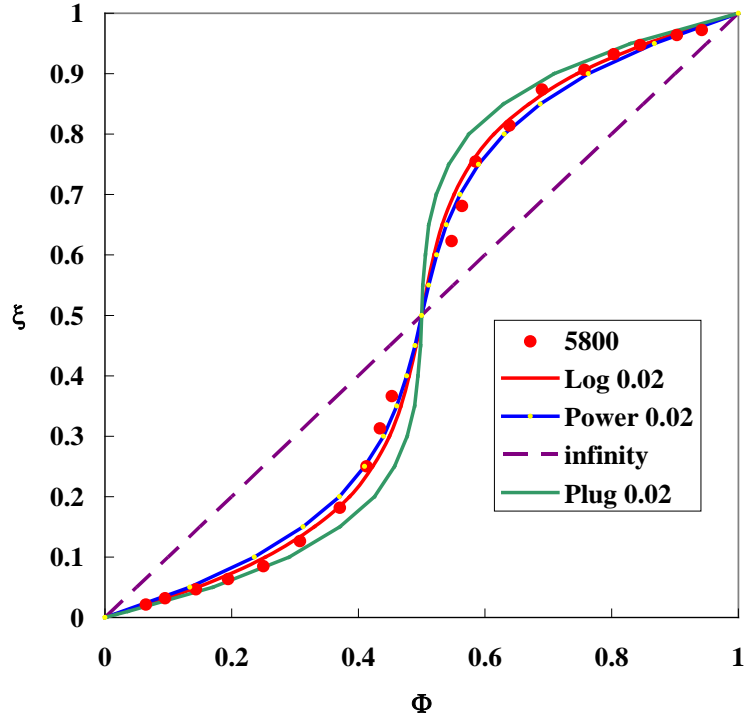

Figure 7. Comparison of velocity profiles predicted from equations (36) and (48) with the measurements of Reichardt (1956) at $Re = 5800$. Numbers indicate values of the time scale $\tau$.

The predictions of equation (48) and (36) have been made with the time scale of $\tau = 0.02$ as predicted by equation (37) and (42) for $Re = 5800$. Variations of the power law exponent $1/11 \leq p \leq 1/8$ did not greatly affect the calculated profiles. The result for $p = 1/9$ is shown in Figure 7. Results for the log-law initial profile discussed in the next section are also shown. Equation (48) fits the data better than equation (36) and fare well near the wall quite well. The greatest difference is found in the core region near the midpoint but the prediction of equation (36) can be improved if we change the value of $\tau$ as shown in Figure 2.

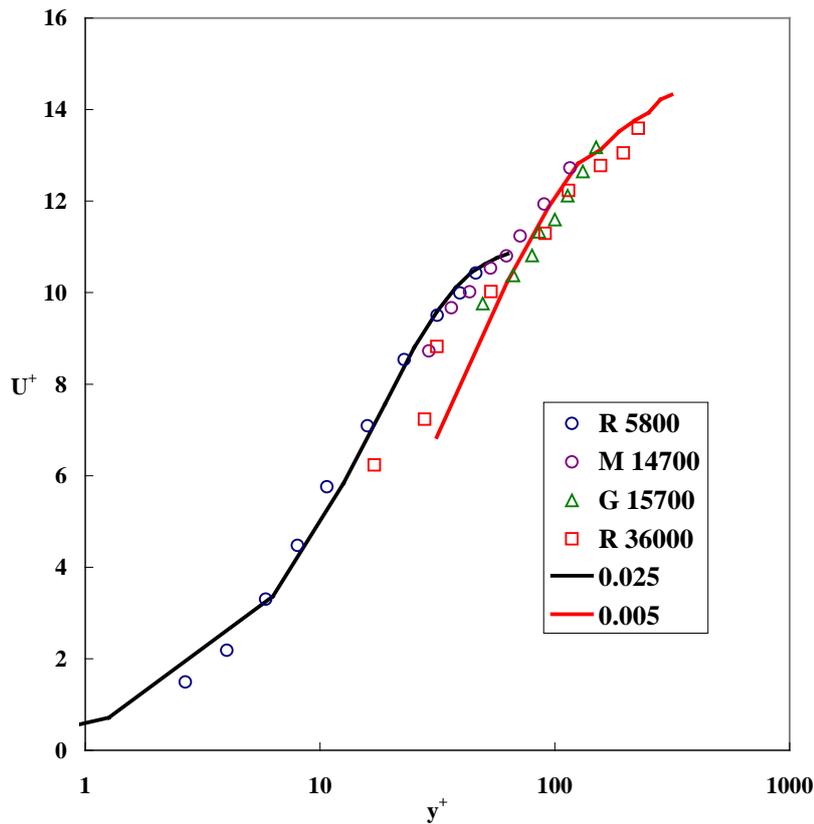

Figure 8   Universal velocity profiles from Reichardt (1956), Geach (1970) and Murphy reported by Geach. Lines are from equation (48) used with an initial power law profile for two time scales.

The semi-log representation that Prandtl called universal velocity profile shows a greater dependence on the Reynolds number than pipe flow. It has been known for some time (Dorfman, 1963) that the in Taylor-Couette flow the ratio of cylinder diameters and curvature of the gap have a non-trivial effect on the shear velocity used to normalise the velocity profile. Nevertheless Figure 7 shows that equation (48) used in conjunction with a power law profile does capture the presence of both a viscous wall layer and a log-law layer. Thus it is not necessary to assume an initial log-law profile to capture the presence of a log-law region in the analysis.

**3.2.2 Initial log-law profile**

An initial log-law profile can be postulated by writing

$$\tilde{u}^+ = 2.5\ln(y^+) + B \tag{49}$$

Equation (49) must pass through the point $\tilde{u}^+ = V^+/2$ at $y^+ = H^+/2$. Hence

$$B = \frac{V^+}{2} - 2.5\ln\left(\frac{H^+}{2}\right) \tag{50}$$

$$\phi_0 = \frac{2.5}{V^+}\ln(2\xi) + 0.5 \tag{51}$$

Where

$$V^+ = \sqrt{\frac{2}{f}} \tag{52}$$

Thus the initial log-law profile is Reynolds number dependent. We also note that the log-law cannot extend to the wall since we cannot have the logarithm of zero. Millikan's analysis always assumed that there is a layer, however thin, where the effects of viscosity predominate. We accommodate this fact by applying equation (51) with singularities for $\xi = 0$ and $\xi = 1$.

While there is little difference between predictions using the power law and log-law initial profiles at low Reynolds numbers, this difference becomes much more pronounced at higher Reynolds numbers as shown in Figure 9 for Re = *36000*. In this case, equation (42) gives $\tau = 0.00085$. The initial plug and power law profiles give very poor predictions unless the time scale $\tau$ is adjusted as shown in Figure 9. The log-law initial profile is the only postulate that gives good predictions with the right time scale. Unfortunately at very small time scales convergence is very slow since sinus is an oscillating function.

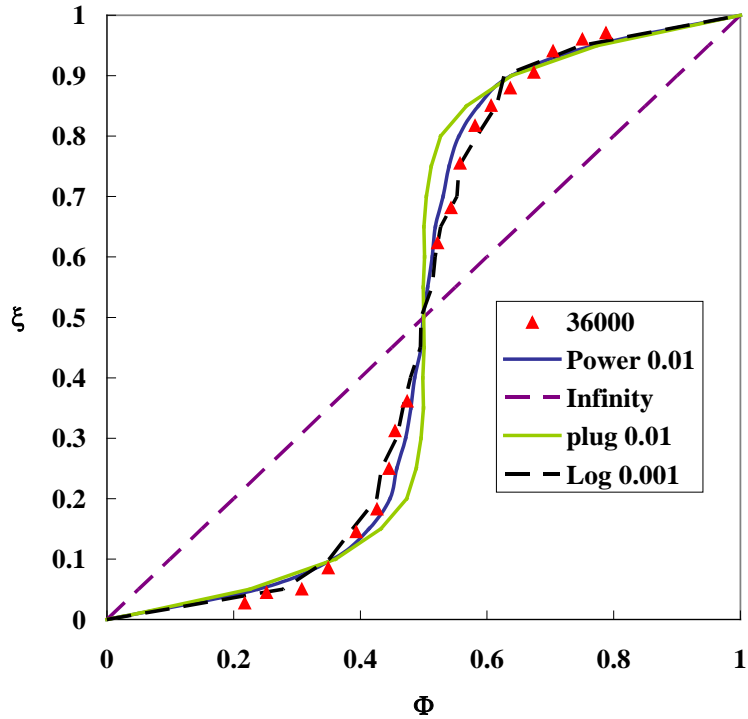

Figure 9 Predicted velocity profiles at Re = *36000*. Data of Reichardt (1956). Numbers indicate values of $\tau$ used.

The original work was done in 1979-1980 in Vietnam with pen, paper and a slide rule, the tools available to the author at the time, and not published till now because contact with foreigners were restricted under the then Stalinist regime. When I emigrated and restarted life for the third time at 50 with five children of school age there was precious little time to spare for pursuits of personal interest. The blue dotted line represents calculations involving the first 18 terms in the series.

## 4   Discussion

The basic physical concept underlying this analysis is that steady state flow can be visualised in terms of an infinite repetition of unsteady state diffusions of momentum from the wall into the mainstream. As such it must be analysed in terms of a Lagrangian derivative along the path of diffusion (Trinh, 2010b). This approach has been shown to accurately obtain all the classical steady state solutions for laminar boundary transport of heat, mass and momentum in both Newtonian and non-Newtonian fluids (1992b, Trinh and Keey, 1992a, Trinh, 2010b). The application of this technique to turbulent flows starting with this paper monitors the evolution of an original profile $\phi_0$ with time as viscous momentum diffuses into the main flow. The time averaging of the instantaneous solutions is based on the principle that low speed streaks of all ages up to a targeted value $\tau$ are detected with equal probability by a fixed probe in the flow field.

The attraction of this approach is that no postulation is made about the many coherent structures that populate turbulent flows as has been extensively documented in the last 50 years. Nor is there any postulation made about the form of the interactions between the different terms in the Navier-Stokes equations in particular no mathematical model is needed for the Reynolds stresses. These statistics can however be extracted using a technique presented previously (Trinh, 2005, 2009c, 2010c) and will be discussed in a separate paper that compares results from different geometries and fluid rheological models.

It is known that flow between rotating cylinders, used in practice to approximate plane Couette flow, involves large scales instabilities first identified by Taylor (1923) that arise from the centrifugal effect not encountered in straight channel flows. The evolution of secondary flows in Taylor-Couette, flow between rotating cylinders, is so complex that is not a very good case study for the general study of turbulence development (Trinh, 2009c). The effect of the large coherent oscillating structures in plane Couette flow discussed in section 1.2 cannot be adressed in this approach.

In fact only two coherent structures are relevant to the physical background underlying this approach: the low speed-streaks where the diffusion of viscous momentum occur as governed by the RHS of equation (10) and the ejections or bursts that interrupt the sweep phase of the low speed streaks by literally taking the wall fluid away triggering an inrush of outer fluid towards the wall. The effect of the intermittent ejections is introduced through the value of the time scale $\tau$ and the initial profile $\phi_0$ but is not modelled specifically in this analysis.

The present analysis is compatible with many approaches of turbulence analysis. There is for example no philosophical difference with the method of proper orthogonal decomposition of velocity traces introduced by Bakewell and Lumley (1967) and recently applied to Taylor-Couette flow by Moehlis et al. (2002) since the series obtained are all orthogonal: Fourier series for Cartesian coordinates (Couette, parallel plates flow), Bessel series for cylindrical coordinates (pipe flow) and Legendre series for spherical coordinates (boundary layers on spheres). Moehlis et al. found that the dominant Eigen mode approximated the mean turbulent velocity profile and attempted to link the next two energetic modes with vortices that are attracted to these fixed points.

We can borrow the terminology of the theory of strange attractors (Langford III, 1982) to describe the essential visualisation in the present approach. Our analysis defines two asymptotic states $\phi_0$ and $\phi_\infty$ that define two basins of attraction corresponding respectively to (laminar) viscous diffusion and full turbulence. The parameter that defines the level of attraction to these alternate states is the normalised time scale $\tau$. Because the viscous dissipation term is always present in the Navier-Stokes equation, there will always a physical space where viscous momentum predominates. This cannot be destroyed but the physical space over which it applies can be reduced with increasing levels of turbulence. This is clearly seen in the instantaneous solutions of equation (10) with decreasing values of $\tau$. In fact when that space is estimated with the use of Stokes' solution for a stream suddenly set in motion the boundaries of this space expressed in normalised terms $\delta_v^+$ and $x_v^+$ this space is remarkably constant once turbulence is triggered (Trinh, 2010a). In that sense what we call the solution of order $\varepsilon^0$ linked to viscous diffusion of momentum (Trinh, 2009c) cannot be destroyed and represents a clear basin of attraction. We note that the profile $\phi_0$ is only relevant once the streaming flow has become strong enough to generate a jet of wall fluid. Thus the ratio of kinetic energy contained in the streaming flow to viscous energy at the point of bursting represented for example by the values of $\delta_v^+$, a local Reynolds number or the corresponding local friction factor $U_v^+$ can be seen as the centre of the attraction basin of turbulence and these parameters can be used effectively to produce truly universal master velocity profiles for all fluids and geometries (Trinh, 2009b, 2010f).

The term orbit in the theory of strange attractors can be best related to the instantaneous velocity profile at each time $\tau$ since Langford refers it to a solution to the differential equation regarded as a curve in the state space. It defines the actual instantaneous profile that is attained at time $\tau$ which also marks the limit of the process of viscous diffusion of momentum.

The scale $\tau$ is itself a function of three parameters, the friction factor $f$, the Reynolds number $\mathrm{Re}$ and the physical time scale $t$ of the process (equation 37). The relation between $f$ and $\mathrm{Re}$ is not fixed for each geometry but will also depend the scale of the disturbances imposed on the flow. It is well known for example that turbulence in boundary layers can be triggered early with a trip wire (Schlichting, 1960) and therefore achieve higher values of $f$ at lower values of $\mathrm{Re}$. Indeed the Reynolds number by itself is not a complete criterion for the onset of turbulence and the solution of order $\varepsilon$ in the study of oscillating boundary layers incorporates the magnitude of the disturbances (fast fluctuations) in the criterion of for development of the streaming flow (Trinh, 2009c). There is continuing interest in the effect of these disturbances, monitored through the intensity of free stream turbulence on the transition to turbulence e.g. (Brandt et al., 2004) and on transport coefficients e.g. (Kondjoyan and Daudin, 1995). Equation (37) clearly indicates that the magnitude of the disturbances required to set up a given level of turbulence (and therefore $\tau$) decreases with increasing Reynolds numbers as has been widely observed e.g. (Hof et al., 2003, Eckhardt et al., 2007a).

The contribution of different Eigen modes to the function $\Phi_t$ is shown in Figure 10 for $\mathrm{Re} = 5800$. There is clearly one dominant Eigen mode corresponding to $n = 2$. The curves in this figure do not represent the instantaneous contributions of the different modes at $\tau = 0.02$ but a statistical average of all contributions of each Eigen mode over an interval $0 \leq \tau \leq 0.02$ assuming they occur with equal probability.

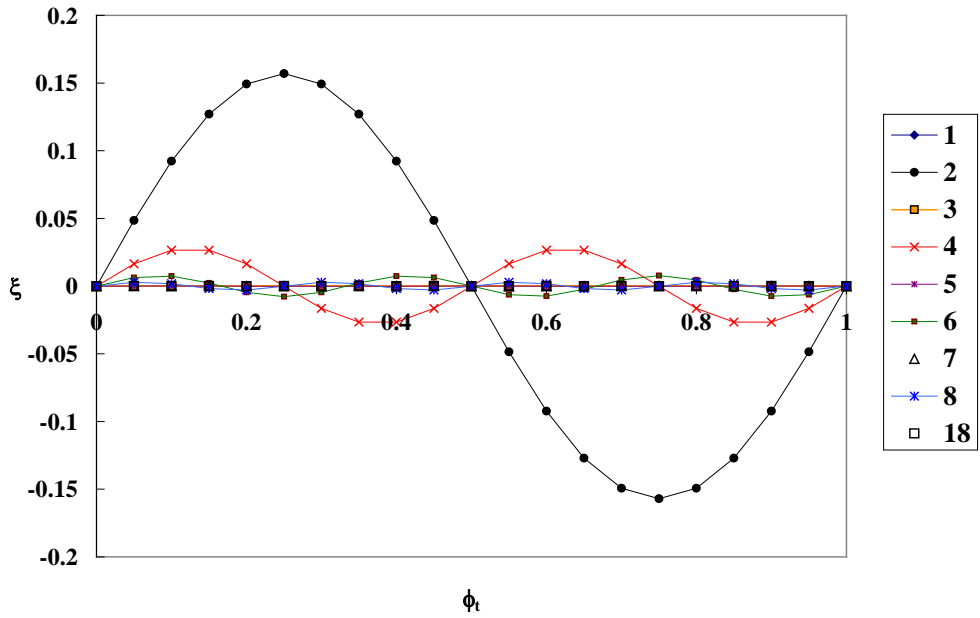

Figure 10 Contributions of the different Eigen modes for $\tau = 0.02$ to the velocity profiles. Initial log-law profile.

This dominant Eigen mode $\Phi_t(n = 2)$ is almost equal to the sum of all contributions, $\Phi_t$ as shown in Figure 11.

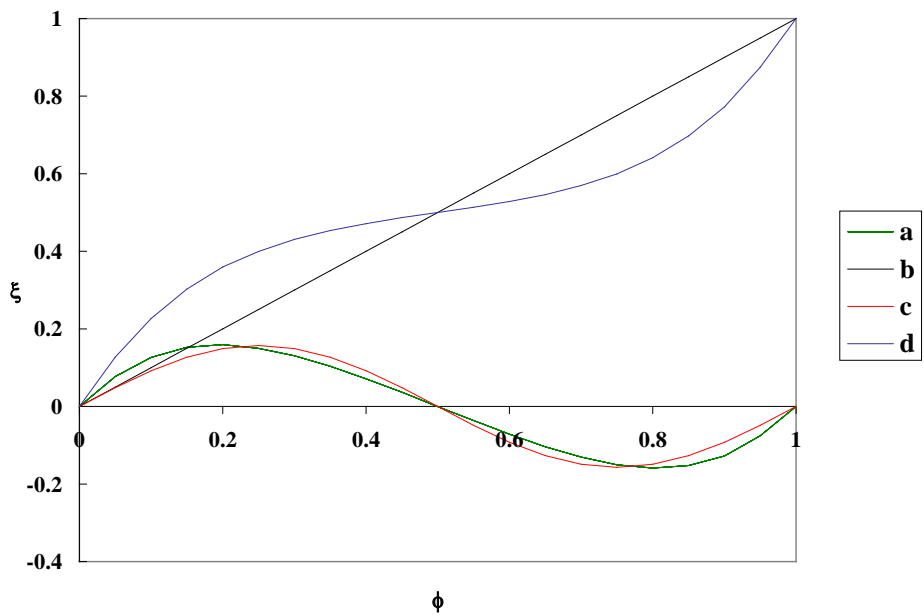

Figure 11 Contributions to the velocity profile at $\tau = 0.02$, Re = $5800$. (a) $\Phi$, (b) $\phi_\infty$, (c) $\Phi_t$, (d) $\Phi_t(n = 2)$.

This is reminiscent of the observations of Bakewell and Lumley (op.cit.) and Moehlis et al. (op.cit.) who also found that the dominant Eigen mode in their proper orthogonal decomposition coincided well with the time averaged profile. The relative contributions of the terms $\phi_\infty$ and $\Phi_t$ to the final profile $\Phi$ are also shown.

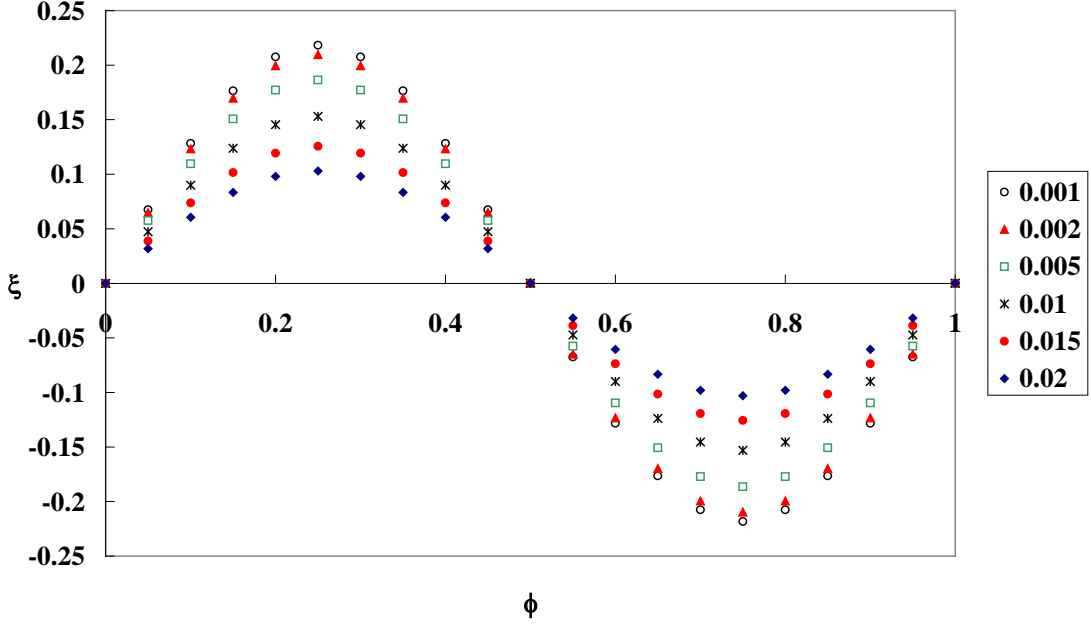

Figure 12  Instantaneous profiles for the Eigen mode $n = 2$, Re $= 5800$, at different times.

However the present solution leads to a physical interpretation of the terms and statistics of the orthogonal series that differ from the classical views. In the statistical theory of turbulence introduced by Taylor (1935) the auto correlation of the instantaneous velocity trace is given by:

$$f(\tau) = \frac{\overline{u(t)u(t+\tau)}}{\overline{u(t)^2}} \qquad (53)$$

where $\tau$ is here a time delay. The two point spatial correlation is

$$f(r) = \frac{\overline{u(x,t)(u(x+r,t)}}{\overline{u(x,t)^2}} \qquad (54)$$

It is used to estimate a macroscale $L_x$ (Bradshaw, 1971) which is traditionally interpreted as the size of the eddy passing the probe.

$$L_x = \int f(r)dr \qquad (55)$$

and a (Taylor) microscale (Bradshaw, 1971, Lesieur, 2008), an estimate of the size of an eddy where viscous dissipation occurs

$$l_t = \frac{15\overline{u'^2}\nu}{\zeta} \tag{56}$$

where $\zeta$ is the energy per unit volume. The energy spectrum describes the flow of energy between different scales and can be obtained from a Fourier transform of the crf

$$\frac{E(k)}{\overline{u'^2}} = \frac{2}{\pi} \int_{-\infty}^{\infty} f(r)\exp(-ikr)dr \tag{57}$$

It is classically interpreted as the description of an energy cascade from the eddies of large size which are inherently unstable to smaller and smaller eddies. In that sense, Moehlis et al. interpret the Eigen modes in the proper orthogonal decomposition to " include pairs and stacks of streamwise rolls and associated streaks".

Since the present solution only deals with the smoothed velocity $\tilde{u}_i$, it cannot capture the travelling vortices that impress fast fluctuations on the low speed streaks and the streaming flow. The Eigen modes in the Fourier series in equations (36) and (48) cannot therefore be interpreted in terms of vortices.

The instantaneous profiles for the dominant Eigen function $n = 2$ are shown in Figure 12. While they follow a typical sine function in space, the evolution over time, shown in Figure 13 for three values of $\xi$, is smooth and non-oscillating since it is governed by the exponential term in equation (448. They cannot capture the effect of the longitudinally vortices observed by Itoh et al. unless an oscillating term is introduced into the initial profile $\phi_0$.

The basic statistics that can be compiled from equation (48) are similar to the statistics obtained from the Stokes solution and do not capture the consequences of the streaming flow (ejections) that are necessary features in turbulent flows. They simply reflect the movement of a front of diffusion of viscous momentum into an initially full turbulent flow field (Trinh, 2010c).

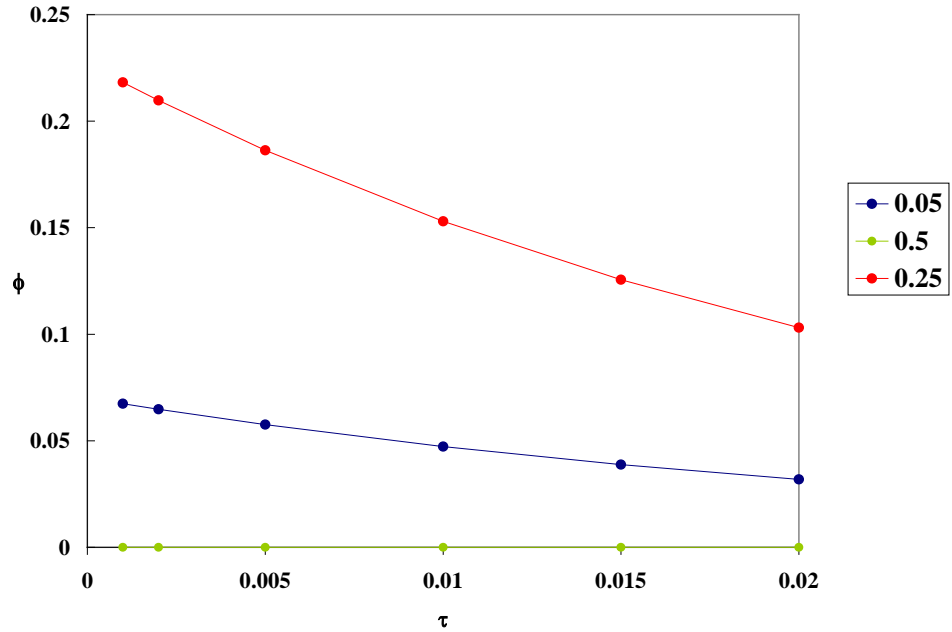

Figure 13  Evolution of the contribution for $n = 2$ at three positions.

# 5    Conclusion

A simple analytical solution for turbulent plane Couette flow has been obtained by modelling the effect of diffusion of viscous momentum from the wall into an initial fully turbulent flow.